\pdfoutput=1
\documentclass[conference]{IEEEtran}
\IEEEoverridecommandlockouts
\usepackage{cite}
\usepackage{amsmath,amssymb,amsfonts}
\usepackage[linesnumbered, ruled, longend]{algorithm2e}
\usepackage{graphicx}
\usepackage{textcomp}
\usepackage{xcolor}
\usepackage{stfloats}
\usepackage{subfigure}
\usepackage{booktabs}
\usepackage{multirow}
\usepackage{enumitem}
\usepackage{fontawesome}
\usepackage{doi}
\usepackage{cite}
\usepackage{hyperref}

\hypersetup{hypertex=blue,
            colorlinks=blue,
            linkcolor=blue,
            anchorcolor=blue,
            citecolor=blue}

\def\BibTeX{{\rm B\kern-.05em{\sc i\kern-.025em b}\kern-.08em
    T\kern-.1667em\lower.7ex\hbox{E}\kern-.125emX}}
\begin{document}

\title{Adversarial Threats to Automatic Modulation Open Set Recognition in Wireless Networks
}

\author{\IEEEauthorblockN{
\textsuperscript{\dag}Yandie Yang,
\textsuperscript{\dag}Sicheng Zhang,
\textsuperscript{\dag}Kuixian Li,
\textsuperscript{\ddag}Qiao Tian,
\textsuperscript{\dag}Yun Lin\textsuperscript{*}}

\IEEEauthorblockA{
\textsuperscript{\dag,*}
\textit{College of Information and Communication Engineering, Harbin Engineering University, Harbin, China} \\
\textsuperscript{\ddag}
\textit{College of Computer Science and Technology, Harbin Engineering University, Harbin, China}\\
\textit{E-mail: \{yangyandie, zhangsc, likuixian\}@ieee.org, tianqheu@163.com, linyun@ieee.org} \\
}
}

\maketitle

\begin{abstract}
Automatic Modulation Open Set Recognition (AMOSR) is a crucial technological approach for cognitive radio communications, wireless spectrum management, and interference monitoring within wireless networks. 
Numerous studies have shown that AMR is highly susceptible to minimal perturbations carefully designed by malicious attackers, leading to misclassification of signals. 
However, the adversarial security issue of AMOSR has not yet been explored. 
This paper adopts the perspective of attackers and proposes an Open Set Adversarial Attack (OSAttack), aiming at investigating the adversarial vulnerabilities of various AMOSR methods. 
Initially, an adversarial threat model for AMOSR scenarios is established. 
Subsequently, by analyzing the decision criteria of both discriminative and generative open set recognition, OSFGSM and OSPGD are proposed to reduce the performance of AMOSR. 
Finally, the influence of OSAttack on AMOSR is evaluated utilizing a range of qualitative and quantitative indicators. 
The results indicate that despite the increased resistance of AMOSR models to conventional interference signals, they remain vulnerable to attacks by adversarial examples.
\end{abstract}


\section{Introduction}
Amidst the swift progress of wireless technology and the contraction of available spectrum resources, the complexity of spectrum sharing, interference diagnostics, and management within wireless networks has escalated significantly~\cite{wei2022toward}.
Automatic Modulation Classification (AMC)~\cite{tu2018semi,9200788} is a crucial step in demodulating signals with unknown modulation types, essential for frequency monitoring and security maintenance in complex wireless networks.
Deep learning (DL) has become a widely used method in AMC~\cite{9128039,9049161}, due to its unique advantages such as autonomous analysis and nonlinear fitting.
Traditional DL-based AMC methods rely on closed set assumptions and fail to account for unknown interference signals intentionally emitted by non-cooperative users, 
thus do not meet the practical needs of real-world environments. 
Consequently, Open Set Recognition (OSR)~\cite{10210314}, which can identify unknown signal types has begun to receive widespread attention.

Given the limitations of deep neural network (DNN) inevitably has a closed set nature due to its use of the traditional softmax cross-entropy loss function during training. 
Bendale et al.~\cite{bendale2016towards} introduce a pioneering solution for open set deep learning architectures by substituting the softmax with an OpenMax that is fine-tuned using the Weibull distribution. 
Wen et al.~\cite{10.1007/978-3-319-46478-7_31} present the CenterLoss function as an innovative supervisory signal for recognition tasks. 
It is designed to ascertain the feature centers of each class and imposes penalties on the divergence between features and their respective class centers.
Kong et al.~\cite{kong2021opengan} synthesizes unknown class samples using Generative Adversarial Networks (GAN) to aid network training, offering explicit probability estimates for the generated unknown class samples. 
Sun et al.~\cite{sun2020conditional} presents an OSR algorithm based on Conditional Gaussian Distribution Learning (CGDL), which utilizes a probabilistic ladder architecture to preserve the representation of input information in intermediate layers and detects unknown samples by approximating latent features to different Gaussian models.

Lately, it has been uncovered by researchers that DNNs are susceptible to being fooled by adversarial examples to produce erroneous outputs.
The concept of adversarial examples is first introduced by Szegedy et al.~\cite{2013arXiv1312.6199S} and it is found that adversarial examples can be transferred between different models.
Lin et al.~\cite{9259112} introduce carefully designed adversarial examples into modulation signals under different environments, validating the potential adversarial threats in AMC. 
To mitigate the frequency leakage and glitch caused by high frequency components in adversarial perturbations, Zhang et al.~\cite{10479476} devised the spectrum-focused frequency adversarial attack algorithm, enhancing the stealthiness of the attacks.
Considering the openness of the electromagnetic space, Qi et al.~\cite{9761960} developed a black box adversarial attack of detection-tolerant that reduces the cost of attacks while increasing the success rate. 
In response to the adversarial threats, researchers have also proposed corresponding defense methods~\cite{10458676,9570781,10418157}.

Although adversarial attacks and defenses are commonly studied in AMC, they have not been extensively explored in AMOSR. 
Rozsa et al.\cite{rozsa2017adversarial} design closed set adversarial attacks against deep features and find that OpenMax is as vulnerable by closed set adversarial perturbation as softmax.
To explore the impact of open set adversarial attacks on AMOSR, this paper proposes Open Set Adversarial Attacks (OSAttack), including Open Set Fast Gradient Attack Methods (OSFGSM) and Open Set Projected Gradient Descent (OSPGD), aiming to draw attention to the adversarial security challenges in OSR.
The main contributions of this paper are as follows:
\begin{itemize}
	\item Building on research in adversarial attacks and AMOSR, we establish an adversarial threat model for AMOSR.
	\item Regarding the discrimination criteria for discriminative and generative models, we propose the attack algorithm of OSFGSM and OSPGD.
	\item We carry out comprehensive experiments to evaluate the performance of AMOSR before and after OSAttack, discovering that AMOSR also exhibits adversarial vulnerabilities.
\end{itemize}
The remainder of this paper is organized as follows:
Section II provides an overview of adversarial attacks and the adversarial threat model of AMOSR.
Section III details the OSAttack algorithm and evaluation indicators.
Section IV discusses the datasets used and experimental validation of OSAttack on effectiveness.
Section V summarizes the findings and suggests directions for future research.
\section{Background and Threat model}\label{sTwo}
\subsection{Background of Adversarial Attack}\label{sThree1}
Adversarial examples $x_{adv}$ are deliberately crafted inputs that aim to fool models by introducing specific perturbations.
FGSM~\cite{2014arXiv1412.6572G} as a classical gradient-based algorithm of adversarial examples generation, which modifies each point in the input sample by an equal amount of increase or decrease with the gradient direction of the loss function for multi classification model \(f_\theta\)
\begin{equation}
	x_{adv}=x+\varepsilon\cdot \text{sign}\big(\nabla_{x}J(\theta,x,y)\big),
\end{equation}
where $\varepsilon$ is the maximum perturbation limit, $\theta$ represents the parameters of \(f_\theta\), $y$ is the label related to input $x$, \(\nabla_x J(\cdot)\) is the gradient of loss function \(J(\cdot)\) with respect $x$, and \(\text{sign}(\cdot)\) represents the sign function. 
PGD~\cite{2017arXiv170606083M} iteratively generates adversarial examples to improve upon FGSM
\begin{equation}
x^{i+1}_{adv}=\prod_{x+\delta}\Big(x^{i}_{adv}+\alpha \text{sign}\big(\nabla_{x}J(\theta,x,y)\big)\Big),
\end{equation}
where \(i\) denotes iteration index, $\delta$ is adversarial perturbation, \(\alpha \) represents the step size of perturbation, and the projection \( \Pi \) assigns its input to the nearest element within the perturbed input set \( x+\delta \).
\subsection{Adversarial Threat Model of AMOSR}\label{sThree2}
The goal of AMOSR is to determine whether $x$ belongs to unknown class and to correctly classify signals of known classes. 
Then, the goal of adversarial attacks against AMOSR is to induce misclassification of unknown signals as known classes by applying subtle perturbations, 
thereby evading electromagnetic spectrum monitoring and interference signal demodulation.

Let the training dataset be $\mathcal{D}_{c}=\left\{ \left( x_i,y_i \right) \right\} _{i=1}^{m}$ in which $x_i\in \mathbb{R}^N$ and $y_i\in \left\{ 1,2,...,K \right\}$, 
and define the potential unknown dataset as $\mathcal{D}_{u}=\left\{\left(x_{u},y_{u}\right)\mid y_{u}\in\{K+1\}\right\}$.
During the testing phase of AMOSR, defining the test dataset $\mathcal{D}_{o}=\left\{\left(x_{o},y_{o}\right)\mid y_{o}\in\{1,2,...,K+1\}\right\}$. 
The open space risk is the ratio of the amount of $x_{u}$ recognized as closed sets to the amount of $x_{o}$ recognized as closed sets
\begin{equation}
R_o\left(D_u,D_o;f_\theta\right)=\frac{\int_{D_u}f_\theta(x)dx}{\int_{D_o}f_\theta(x)dx}.
\end{equation}

It can be observed that an increased misclassification of more \(x_u\) as known signals will raise \(R_o\),
therefore, the objective of adversarial attacks can be expressed as
\begin{equation}
	\begin{aligned}
		\begin{aligned}
			\max_{\delta} &R_o\left(D_{u,adv},D_{o,adv};f_\theta\right), \\
			\text{s.t. } &f_\theta(x_{u,adv}) \neq K+1, \\
			&x_{u,adv} = x_u + \delta \in \mathcal{D}_{u,adv},
	    \end{aligned}
	\end{aligned}
\end{equation}
where $\mathcal{D}_{u,adv}$ is the unknown dataset after adding adversarial examples,
$\mathcal{D}_{o,adv}$ is the test dataset that includes adversarial unknown data.
we aim increase \(R_o\) of target AMOSR model by adding minor perturbations to the received unknown signal.
\section{Attack Methods and Measurement Indicator}\label{sThree}
This section describes the proposed OSAttack methods and attack performance measures. 
OSFGSM and OSPGD are developed for the discriminative and generative models.
\subsection{OSAttack of Discriminative Model}\label{sThree1}
\subsubsection{\bf OpenMax and CenterLoss}
Discriminative models usually classify by increasing the difference between closed and open sets and setting thresholds $\tau$. 
We analyze the discriminative rules of OpenMax, CenterLoss, and CGDL, 
and propose corresponding methods for generating adversarial examples.
\begin{figure}[htpb]
	\centering
	\includegraphics[width=1.0\linewidth]{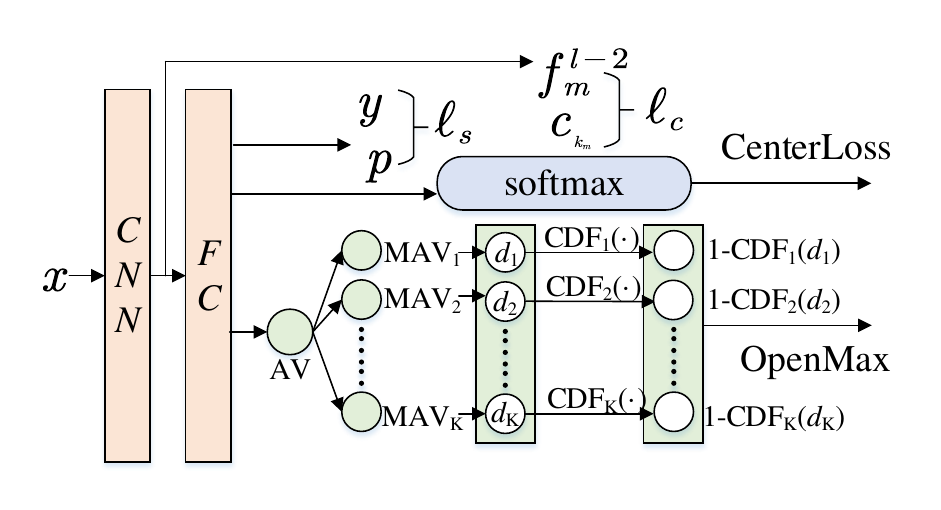}
	\caption{The OSAttack framework diagram for OpenMax and CenterLoss.~~~~~~~~~~~}
	\label{Fig_OSAttack-OC}
\end{figure}

As illustrated in Fig.~\ref{Fig_OSAttack-OC}, OpenMax method initially trains through DNN with by minimizing \(\ell_s \)
\begin{equation}
	\ell_s=CE\left(p,y\right),
\end{equation}
where $p$ is the logits, $CE\left(\cdot,\cdot\right)$ represents cross-entropy loss function.
It computes Mean Activation Vectors (MAVs) for each class from correctly recognized samples, and fits a Weibull distribution for each class using the distance between these samples and their MAVs.

During the testing phase, the algorithm calculates the distance \(d\) between the activation vector (AV) and the MAV of each class, obtaining probabilities \( CDF(d) \) through the $CDF\left( \cdot \right)$ function. 
After adjusting weights with \( 1 - CDF(d) \), it uses softmax to compute the new class probabilities $P\left( y=k|x \right)$.
Define $y^*=\operatorname{argmax}P\left( y=k|x \right)$, represents the predicted label.
when \(y^* = K+1\) or $P\left( y=y^*|x \right)<\tau$, $x$ will be classified as belonging to an unknown class.

The CenterLoss method adds a loss function \(\ell_c\) on the basis of \(\ell_s\)
\begin{equation}
\ell_c=\frac{1}{2}\sum_{m=1}^M\|f_{m}^{l-2}-c_{k_m}\|_2^2,
\end{equation}
where $f_{m}^{l-2}$ represents the features before the fully connected layer, \( c_{k_m} \) is the center for the deep feature corresponding to the \( k \)-th class of input samples.
This approach reduces intra-class distances while enlarging inter-class distances, effectively distinguishing between classes.
Finally the predicted probability is output by softmax, when $P\left( y=y^*|x \right)<\tau$, \( x \) is determined by CenterLoss to come from an unknown class.

To sum up, OpenMax recalibrates AVs based on Weibull distribution fitting and CenterLoss utilizes two loss functions training approach, 
thereby minimizing the open space for each known class and enhancing the probability distinction between open set and closed set classes. 
Accordingly, our attack aims to increase the probability that open set samples are misclassified as the most likely closed set classes, rendering open and closed sets more indistinguishable.
Therefore, the adversarial examples generated by the OSFGSM and OSPGD algorithms are respectively denoted as \eqref{OSFGSM1} and \eqref{OSPGD1}:
\begin{equation}
	x_{u,adv}=x_u-\varepsilon\text{sign}(\nabla_{x_u}CE\left(p_u,y^* \right)),
	\label{OSFGSM1}
\end{equation}
\begin{equation}
	x_{u,adv}^{i+1}=\prod_{x+\delta}\left(x_{u,adv}^i-\alpha \text{sign}(\nabla_{x_u}CE\left(p_u,y^* \right)) \right).
	\label{OSPGD1}
\end{equation}
\subsubsection{\bf CGDL}
\begin{figure}[htpb]
	\centering
	\includegraphics[width=1.0\linewidth]{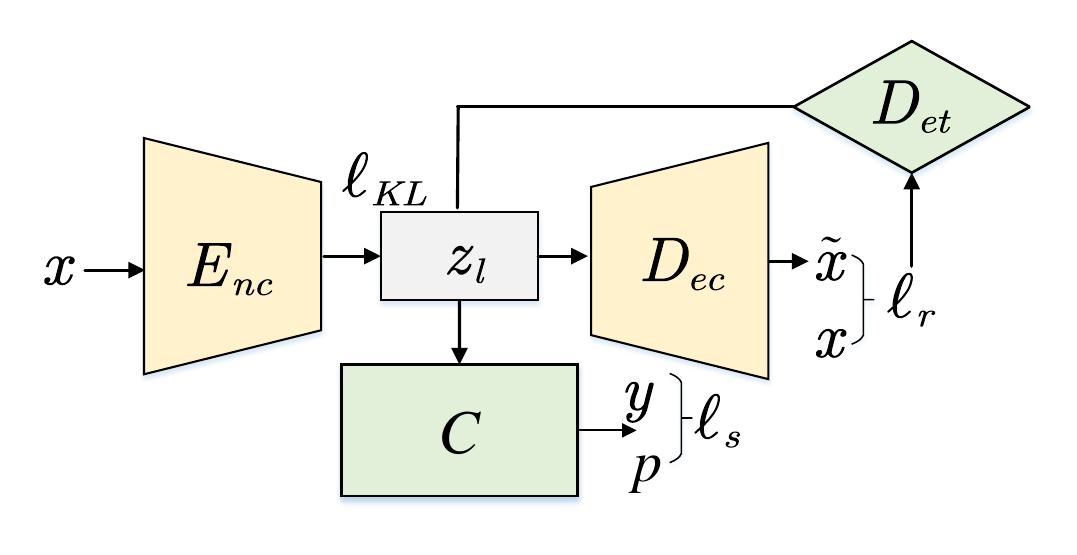}
	\caption{The OSAttack framework diagram for CGDL.~~~~~~~~~~~~~~~~~~~~~~~~~~~~~~~~~~~~~~~~~~~~}
	\label{Fig_OSAttack-CGDL}
\end{figure}

The framework diagram of the CGDL method is shown in Fig.~\ref{Fig_OSAttack-CGDL}, which utilizes a variational autoencoder to construct reconstructed features, and classifies them by enforcing various Gaussian models approximated through different latent features.

The encoder \(E_{nc}\) and decoder \(D{ec}\) extract abstract latent features $z_l$ by using a probabilistic ladder architecture, which in consequence achieves the desired reconstructed signal \(\tilde{x}\)
\begin{equation}
z_l=E_{nc}\left( x \right),
\end{equation}
\begin{equation}
\tilde{x}=D_{ec}\left( z_l \right).
\end{equation}

Then, $z_l$ is used as input to the classifier $C$ to get the predicted labels of the known samples
\begin{equation}
y^*=\operatorname{argmax}\left( C\left( z_l \right) \right).
\end{equation}

The training objective of CGDL is to minimize the sum of reconstruction loss \(\ell_r\), KL divergence \(\ell_{KL}\) and classification loss \(\ell_s\).
Specifically, the mean square error $MSE\left(x,\tilde{x}\right)$ is used to calculate \(\ell_r\), the $CE\left(p,y\right)$ is used to calculate \(\ell_s\) and \(\ell_{KL}\) is calculated in the latent space and middle layers. 
In the testing phase, the unknown detector \(D_{et}\) will determine whether a sample is unknown by modeling the $z_l$ and \(\ell_r\).

By combining \(\ell_r\) and $z_l$ for double discrimination, the CGDL method effectively reduces the open set risk of the input signal.
In this paper, \(\ell_r\) and \(\ell_s\) constitute the constrain of the OSAttack used to generate the adversarial example, aiming to decrease the reconstruction error of open set signals by reducing \(\ell_r\) and to increase the probability that open set signals are classified as closed set signals by encoder $E_{nc}$ and classifier $C$ through the reduction of \(\ell_s\), minimizing the difference between the $z_l$ of open and closed set. 
Therefore, the loss function of OSAttack generating adversarial example for CGDL model is
\begin{equation}
\ell_{CGDL}=\frac{1}{2}CE\left(p_u,y^*\right) +\frac{1}{2}MSE\left(x_u,\tilde{x}_u\right),
\end{equation}
consequently, the OSFGSM and OSPGD algorithms for generating adversarial examples can be represented respectively as in \eqref{OSFGSM2} and \eqref{OSPGD2}
\begin{equation}
x_{u,adv}=x_u-\varepsilon\text{sign}(\nabla_{x_u}\ell_{CGDL}),
\label{OSFGSM2}
\end{equation}
\begin{equation}
x_{u,adv}^{i+1}=\prod_{x+\delta} \left( x_{u,adv}^i - \alpha \text{sign}(\nabla_{x_u}\ell_{CGDL}) \right).
\label{OSPGD2}
\end{equation}
\subsection{OSAttack of Generative Model}\label{sThree2}
\begin{figure}[htpb]
	\centering
	\includegraphics[width=1.0\linewidth]{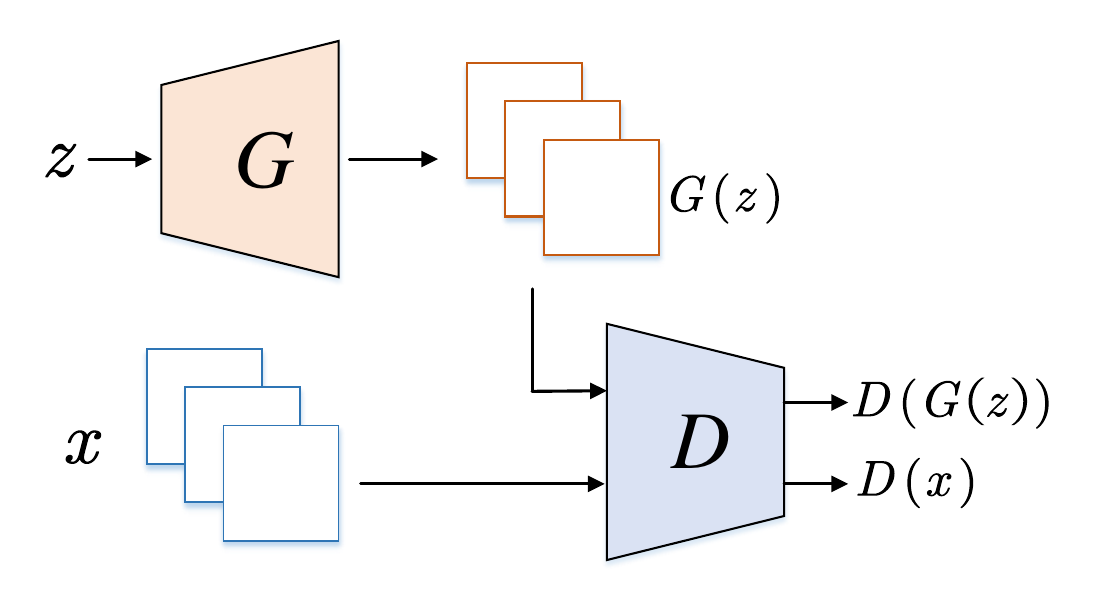}
	\caption{The OSAttack framework diagram for OpenGAN.~~~~~~~~~~~~~~~~~~~~~~~~~~~~~~~~~~~~~}
	\label{Fig_OSAttack-OpenGAN}
\end{figure}
In generative models, it is common for GANs to be employed to generate unknown samples so that DNNs can be exposed to a multitude of unknown samples.
In Fig.~\ref{Fig_OSAttack-OpenGAN}, OpenGAN is a method that utilizes GANs to enhance OSR performance, 
where the discriminator $D$ aims to distinguish between real and fake samples, while the generator $G$ is used to produce fake samples.

First, $D$ is optimized to achieve good discriminative performance
\begin{equation}
\underset{D}{\max}\mathbb{E}_{x\sim \mathcal{D}_{tr}}[\log D(x)]+\mathbb{E}_{z\sim p(z)}[\log(1-D(G(z)))],
\end{equation}
where $z$ as random noise taken from a Gaussian distribution $p(z)$, 
and $G(z)$ represents the fake samples produced by the generator.

To produce more realistic samples, it is necessary to optimize the \(G\) so that the samples it generates can effectively deceive the \(D\)
\begin{equation}
\underset{G}{\min}-\mathbb{E}_{z\sim p(z)}[\log D(G(z))].
\end{equation}

It is the overall training goal of OpenGAN to optimize both loss functions, 
so that the \(G\) learns to generate more and more real samples, and the $D$ is able to distinguish between real and fake samples more effectively.
Thus, this paper attacks the output of $D$ within this structure, 
increasing the error with which the $D$ incorrectly identifies open set samples
\begin{equation}
	\ell_{OpenGAN}=BCE\left(D(x),y_g \right),
\end{equation}
where $y_\text{g}=D\left(x\right) >\tau_\text{g} $, $\tau_\text{g}$ representing the probability threshold, which is constant close to 1. 
corresponding the OSFGSM and OSPGD method for generating adversarial examples can be expressed as \eqref{OSFGSM3} and \eqref{OSPGD3}
\begin{equation}
x_{u,adv}=x_u+\varepsilon\text{sign}(\nabla_{x_u}\ell_{OpenGAN}),
\label{OSFGSM3}
\end{equation}
\begin{equation}
x_{u,adv}^{i+1}=\prod_{x+\delta}\left(x_{u,adv}^i+\alpha \text{sign}(\nabla_{x_u}\ell_{OpenGAN})\right).
\label{OSPGD3}
\end{equation}
\subsection{Evaluation of Indicators}\label{sThree3}
Considering previous studies have widely proven that adversarial attacks pose a significant threat to CSR, 
this paper will not discuss it in detail but focuses primarily on the ability of AMOSR to recognize unknown class.
In order to analyze the threat of adversarial attacks against AMOSR, we choose the accuracy of unknown samples (AUS)~\cite{mendes2017nearest} as the evaluation indicator
\begin{equation}
\mathrm{AUS}=\frac{\mathrm{TU}}{\mathrm{TU}+\mathrm{FU}},
\end{equation}
where $\mathrm{TU}$ and $\mathrm{FU}$ are the number of correctly and incorrectly recognized unknown class samples, respectively.

In addition, we evaluate the performance of AMOSR using the Area Under the Receiver Operating Characteristic curve (AUROC). 
AUROC is the area under the ROC curve, which consists of True Positive Rate (TPR) and False Positive Rate (FPR).
\begin{align}
	\mathrm{TPR}=\frac{\mathrm{TP}}{\mathrm{TP}+\mathrm{FN}},
\end{align}
\begin{align}
	\mathrm{FPR}=\frac{\mathrm{FP}}{\mathrm{FP}+\mathrm{TN}},
\end{align}
where $\mathrm{TP}$ is the amount of true positives, $\mathrm{FN}$ is the amount of false negatives, $\mathrm{FP}$ is the amount of false positives, and $\mathrm{TN}$ is the amount of true negatives. 
The value of AUROC ranges from 0 to 1, which assesses the ability of model to distinguish between two classes by comparing TPR and FPR at various thresholds. 
This metric does not rely on specific classification thresholds, making it particularly suitable for evaluating model performance across different thresholds.
\section{Experiments and Discussions}\label{sFour}
In this section, We describe a detailed introduction to the datasets used and conduct extensive experiments to validate the OSFGSM and OSPGD algorithms.
\begin{figure}[htbp]
	\centering
	\subfigure[\scriptsize OSFGSM]{
		\includegraphics[width=0.45\linewidth]{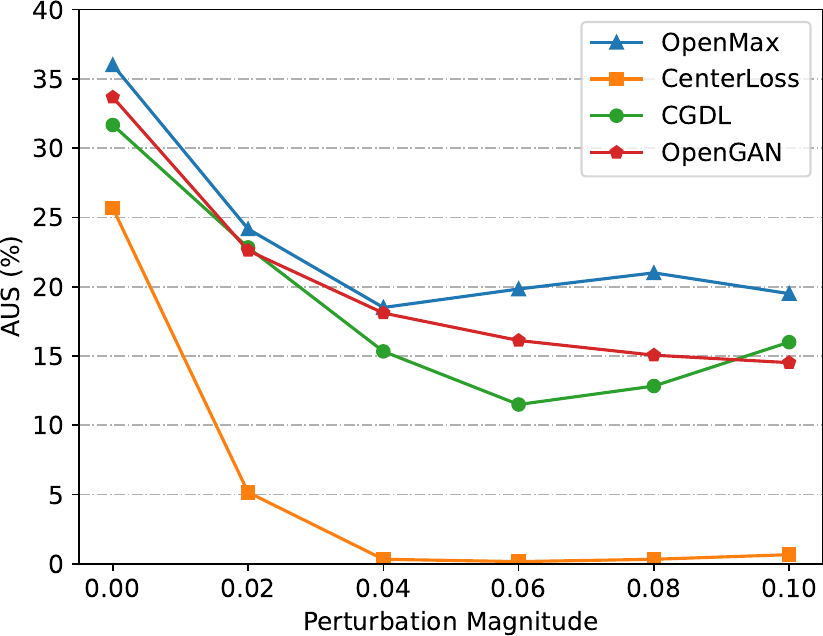}
		\label{osra1}}
	\subfigure[\scriptsize OSPGD]{
		\includegraphics[width=0.45\linewidth]{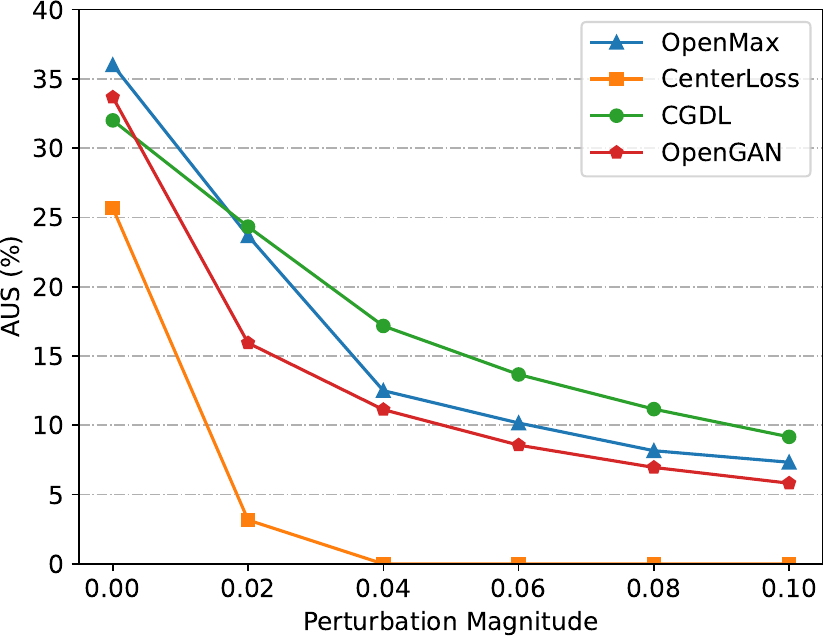}
		\label{osra2}}
	\caption{The AUS of different AMOSR models under OSFGSM and OSPGD attacks as the perturbation changes.}
	\label{Fig_osra}
\end{figure}
\begin{figure}[htbp]
	\centering
	\subfigure[\scriptsize OSFGSM]{
		\includegraphics[width=0.46\linewidth]{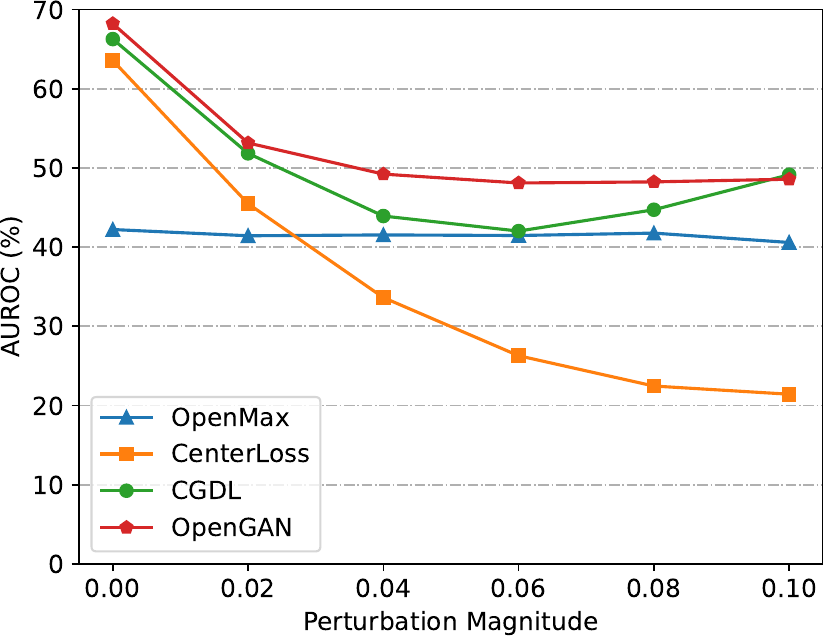}
		\label{auroc1}}
	\subfigure[\scriptsize OSPGD]{
		\includegraphics[width=0.46\linewidth]{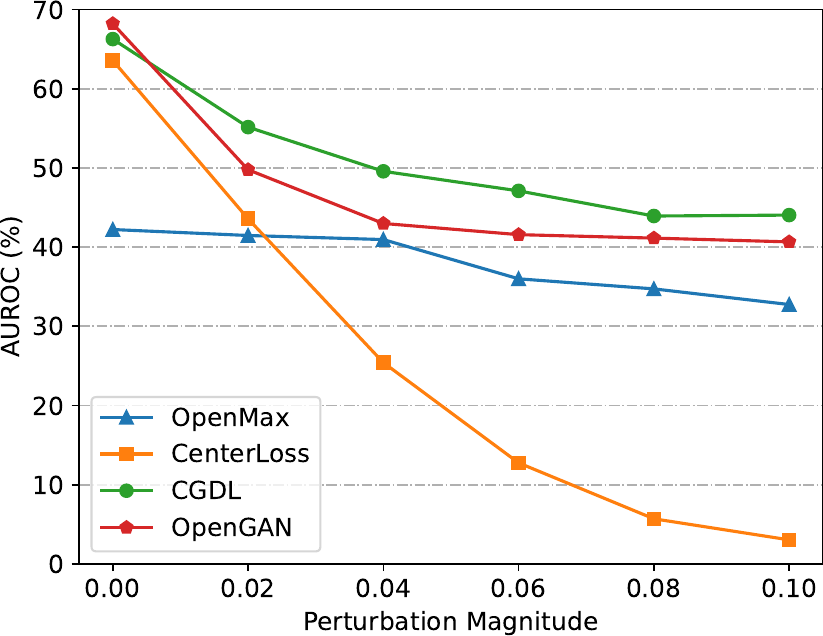}
		\label{auroc2}}
	\caption{The AUROC of different AMOSR models under OSFGSM and OSPGD attacks as the perturbation changes.}
	\label{Fig_auroc}
\end{figure}
\subsection{Dataset}\label{sFour1}
To investigate adversarial examples in wireless communication and their impact on AMOSR systems, 
this paper utilized the open-source simulated dataset RADIOML2016.10A developed by DEEPSIG. 
This publicly available dataset contains eight digitally modulated signals and three analog signals with varying signal-to-noise ratios(SNR). 
In our AMOSR task, eight categories were designated as known classes (BPSK, GFSK, CPFSK, PAM4, QAM64, WBFM, AM-DSB, AM-SSB) and three as unknown classes (8PSK, QPSK, QAM16). 
The dataset consists of 220,000 samples across 20 SNR levels (from -20dB to 18dB) with 2,000 samples for each signal type. 
In this work, it selects 18dB signals for experiments and divides the dataset into training, validation and test sets in the ratio of 6:2:2. 
For each signal vector, it consists of an in-phase component and an quadrature component, each of which has a length of 128.
\begin{figure*}[htbp]
	\centering
	\subfigure[\scriptsize OpenMax]{
		\includegraphics[width=0.23\linewidth]{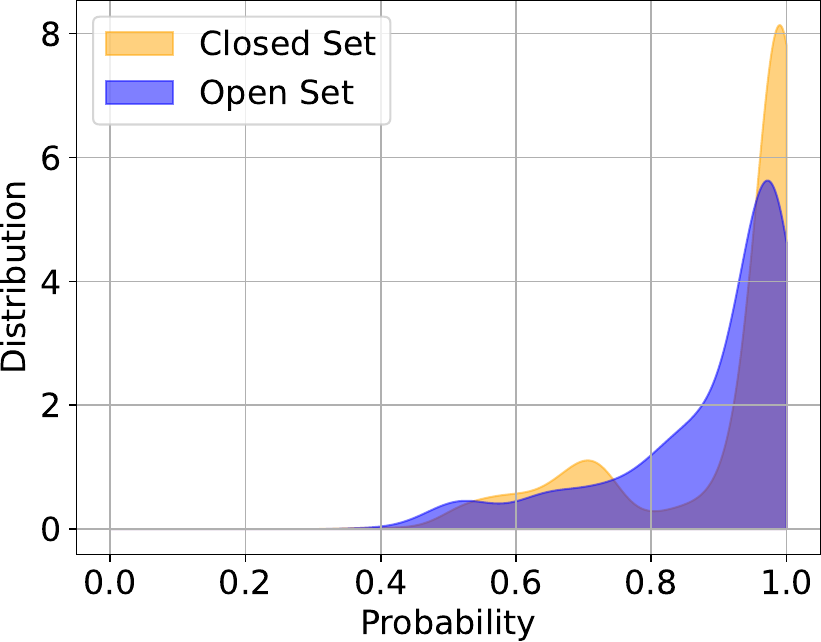}
		\label{distribution1}}
	\hfill
	\subfigure[\scriptsize CenterLoss]{
		\includegraphics[width=0.23\linewidth]{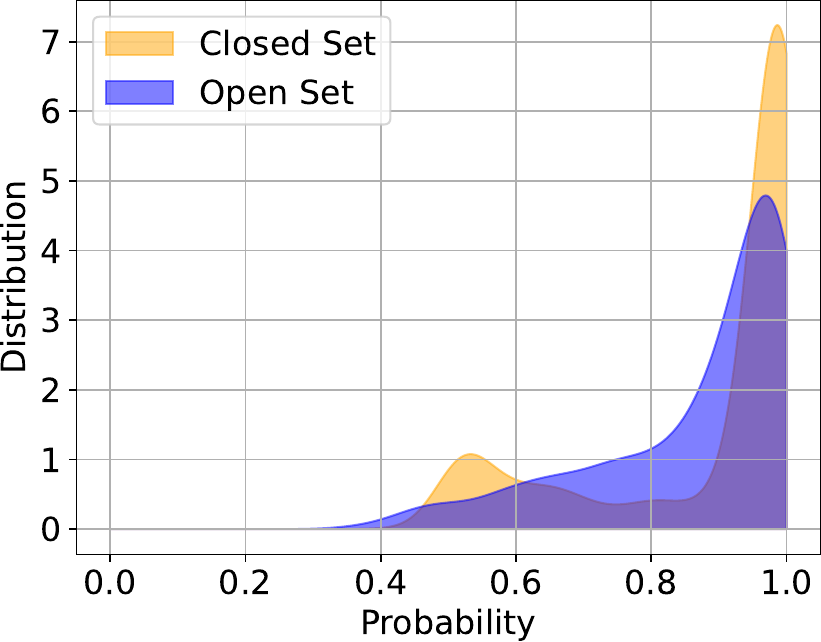}
		\label{distribution2}}
	\hfill
	\subfigure[\scriptsize CGDL]{
		\includegraphics[width=0.23\linewidth]{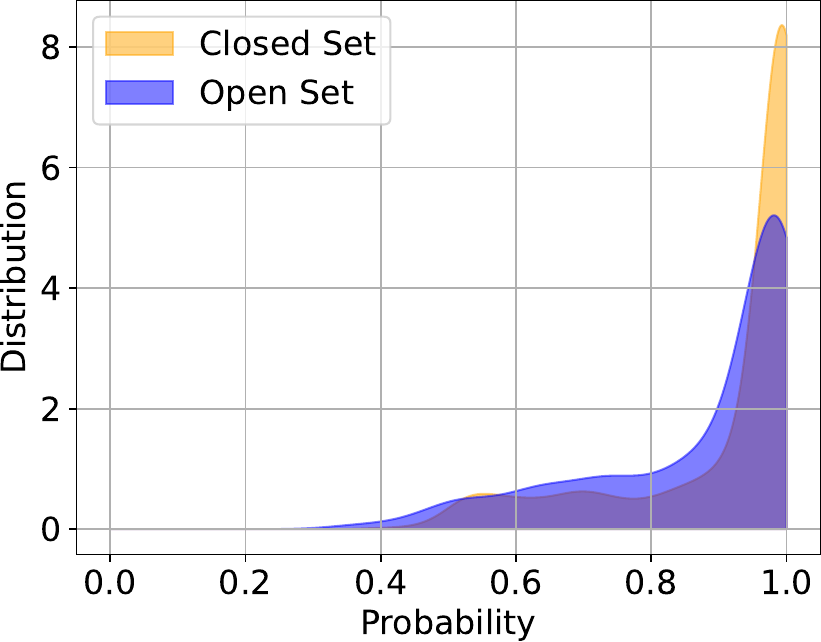}
		\label{distribution3}}
	\hfill
	\subfigure[\scriptsize OpenGAN]{
		\includegraphics[width=0.23\linewidth]{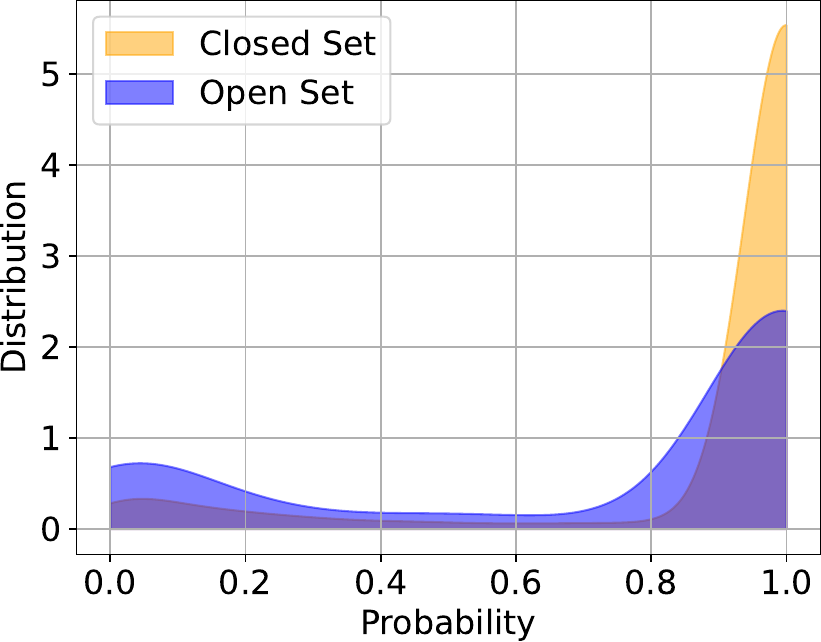}
		\label{distribution4}}
	
	\subfigure[\scriptsize OpenMax]{
		\includegraphics[width=0.23\linewidth]{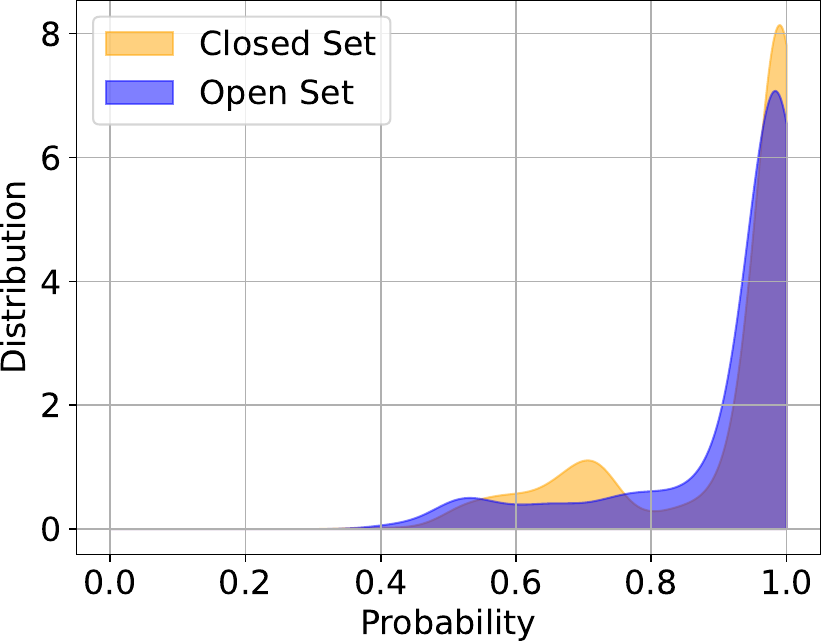}
		\label{distribution5}}
	\hfill
	\subfigure[\scriptsize CenterLoss]{
		\includegraphics[width=0.23\linewidth]{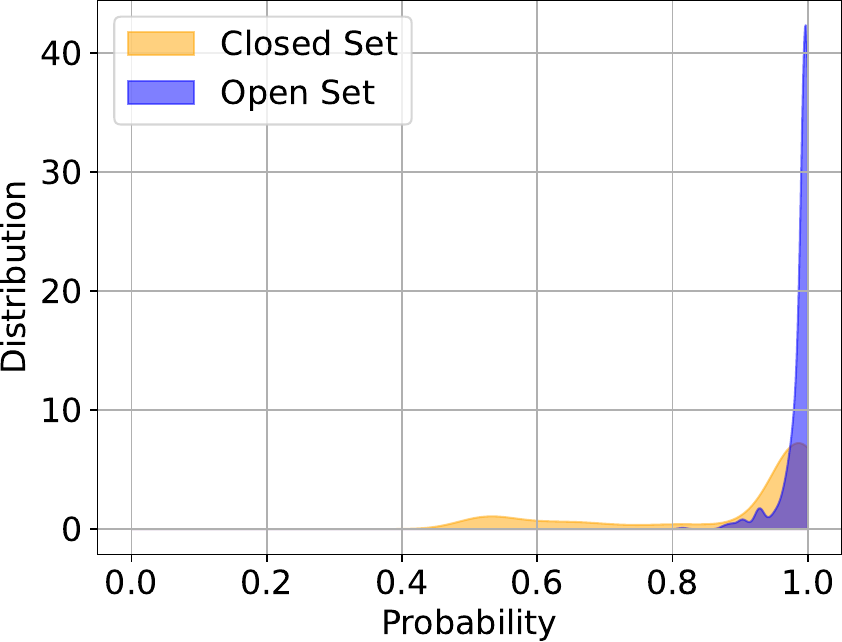}
		\label{distribution6}}
	\hfill
	\subfigure[\scriptsize CGDL]{
		\includegraphics[width=0.23\linewidth]{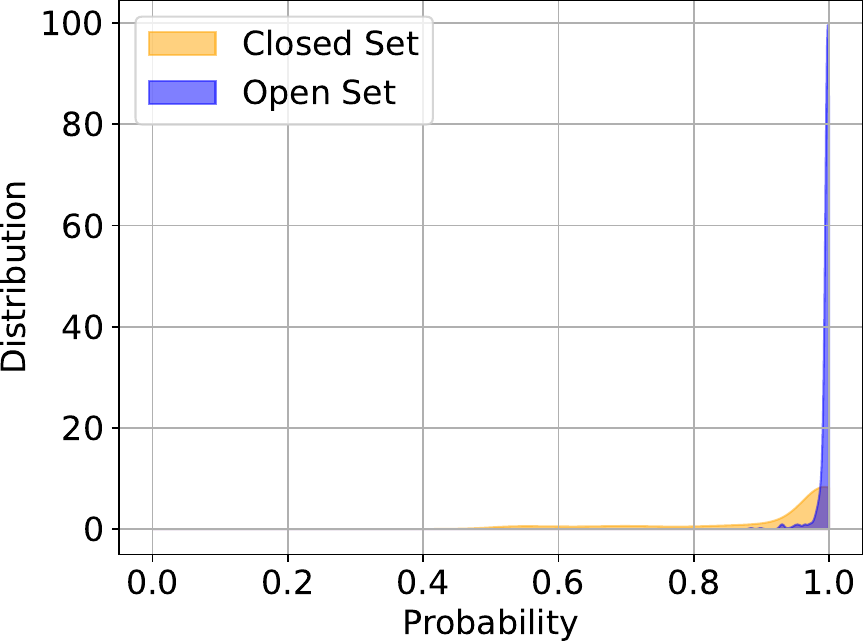}
		\label{distribution7}}
	\hfill
	\subfigure[\scriptsize OpenGAN]{
		\includegraphics[width=0.23\linewidth]{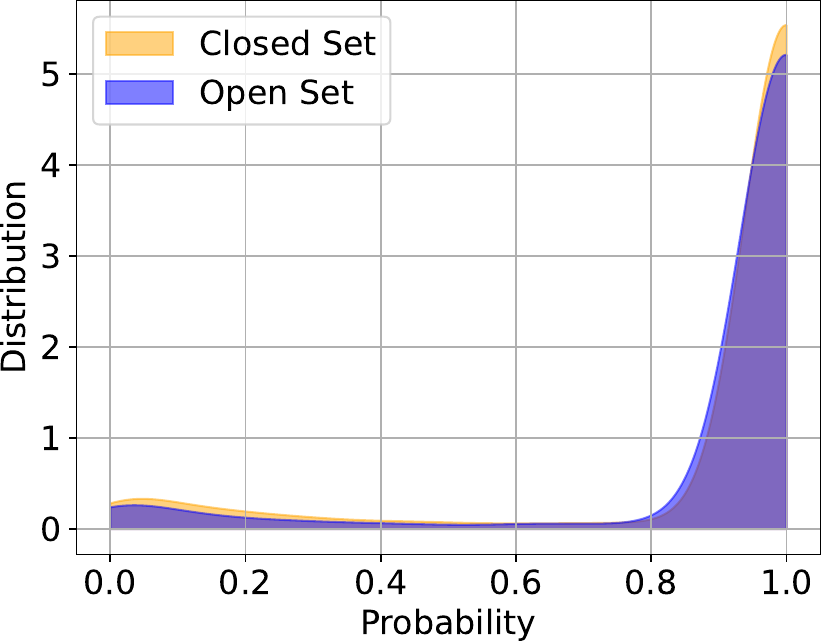}
		\label{distribution8}}
	\caption{Comparison of probability distribution in between closed set and open set.}
	\label{Fig_distribution}
\end{figure*}
\begin{table*}[htpb]
	\centering
	\caption{Comparison of AUS across different OSR methods under various OSAttack.}
	\begin{tabular}{@{}llllllllll@{}}
		\toprule
		&                                                            & \multicolumn{2}{l}{OpenMax(\%)}                                                                           & \multicolumn{2}{l}{CenterLoss(\%)}                                                                            & \multicolumn{2}{l}{CGDL(\%)}                                                                            & \multicolumn{2}{l}{OpenGAN(\%)}                                                        \\ \cmidrule(l){3-10}
		\multirow{-2}{*}{Model} & \multirow{-2}{*}{Clean(\%)} & OSFGSM                             & OSPGD                                                  & OSFGSM                             & OSPGD                                                  & OSFGSM                             & OSPGD                                                  & OSFGSM                             & OSPGD                             \\ \midrule
		OpenMax         & \multicolumn{1}{l|}{23.33}             & {\textbf{4.50}}        &\multicolumn{1}{l|}{\textbf{1.33}}        & 18.33                                                 & \multicolumn{1}{l|}{16.67}                                           & 28.50                                                 & \multicolumn{1}{l|}{18.33}                                           & 25.83                                                  & 26.33\\
		CenterLoss      & \multicolumn{1}{l|}{25.67}            & 20.5                     & \multicolumn{1}{l|}{15.83}              & {\textbf{0.17}}                                       & \multicolumn{1}{l|}{\textbf{0.0}}                                 & 15.0                                                 & \multicolumn{1}{l|}{16.83}                                         & 28.0                                                  & 26.33                                    \\
		CGDL            & \multicolumn{1}{l|}{33.33}            & 69.83                       & \multicolumn{1}{l|}{65.33}               & 65.17                                              & \multicolumn{1}{l|}{56.0}                                            & {\textbf{11.5}}                                   & \multicolumn{1}{l|}{\textbf{13.67}}                                   & 74.33                                              & 73.67                                 \\
		OpenGAN         & \multicolumn{1}{l|}{33.67}         & 36.0              & \multicolumn{1}{l|}{37.5}                         & 35.5                                               & \multicolumn{1}{l|}{39.0}                                             & 6.0                                              & \multicolumn{1}{l|}{1.0}                                            & {\textbf{16.89}}                                   & {\textbf{8.57}}                                 \\ \bottomrule
	\end{tabular}
	\label{Tb_OSRA}
\end{table*}
\subsection{Analysis of AUS}\label{sFour2}
We explore the trends of the AUS of various OSR algorithms under different perturbation magnitudes after being attacked by OSFGSM and OSPGD in Fig.~\ref{Fig_osra}(a) and Fig.~\ref{Fig_osra}(b). 
It can be observed that the AUS of four methods decreases significantly after the attack, with the CenterLoss algorithm experiencing the most pronounced decrease.
In the case of the OSFGSM attack, a mere 0.04 perturbation is sufficient to reduce the AUS of CenterLoss to nearly zero. Similarly, in the OSPGD attack, even slight perturbations lead to a sharp decrease in AUS.

Additionally, for the OSFGSM attack, as the intensity of the attack increases, the AUS of the OpenMax and CGDL algorithms initially shows a gradual decline. 
However, when the perturbation exceeds 0.06, the AUS begins to gradually increase. This phenomenon is not observed in the OSPGD attacks. 
The difference primarily stems from the single step nature of the OSFGSM attack, which relies on a substantial gradient update to generate adversarial examples. 
In contrast, PGD finely tunes the perturbations through continuous multi step iterations, effectively preventing the model from correctly recognizing the samples due to excessive perturbation, thereby maintaining the efficacy and stability of the attack.
\subsection{Analysis of AUROC}\label{sFour3}
The change in AUROC with perturbation magnitudes is investigated in Fig.~\ref{Fig_auroc},
as the perturbation magnitude increases, the AUROC also displays a gradual decline.
Specifically, when the perturbation magnitude reaches 0.06, the AUROC for CenterLoss, openGAN, and CGDL all decline by more than 15\%, dropping from initially high levels to below 50\%. 
In contrast, the decline in AUROC for OpenMax following an attack is less pronounced compared to the aforementioned methods, possibly because OpenMax inherently has a lower AUROC. 
Under the more elaborate OSPGD attack, there is still a substantial drop in performance for OpenMax, which drops by about 10\%.
The above results show that the OSAttack algorithm based on AMOSR decision criteria proposed in this paper significantly weakens the ability of the model to reject unknown samples, 
which reveals that adversarial attacks pose a wide range of threats to the AMOSR system and may lead to poorer recognition performance.
\subsection{Analysis of Probability Distribution}\label{sFour4}
In order to show in detail the changes in the output of the AMOSR model after adversarial attacks,
we present the maximum predictive probability distributions of OpenMax, CenterLoss, and CGDL, as well as the predictive probability distribution of OpenGAN for real samples. 
Fig.~\ref{Fig_distribution}(a)-(d) and Fig.~\ref{Fig_distribution}(e)-(g) respectively demonstrate the probability distribution before and after OSAttack.
In this part of the experiment, it is used the perturbation magnitude of 0.06.

It can be seen that the probability distribution of closed set signals before the model is attacked by OSAttack is mainly concentrated in the higher probability region, while the distribution of open set signals is more dispersed. 
However, after the attack, the predicted probability of the AMOSR model for open set data increases, migrates to high probability regions, and overlaps significantly with the distribution of closed set signals, which increases the difficulty of threshold-based classification discrimination.
Particularly, CenterLoss and CGDL predict probabilities close to 1 for most open set signals, demonstrating that even high confidence levels are no longer reliable for distinguishing unknown class.
\subsection{Ablation Experiment}\label{sFour4}
We analyze the effectiveness of the adversarial examples generated by OSAttack on the AMOSR model using an adversarial perturbation magnitude of 0.06 against other models in Table \ref{Tb_OSRA}, 
where the rows represent the models used to generate adversarial perturbations, and the columns indicate the models used for testing AUS. 
The diagonally bolded data reflects the AUS of a white box attack, while the other data shows the AUS of a black box attack.
Results show that there is a significantly more effective white box attack, and adversarial examples generated using CGDL exhibit the best transferability, effectively deceiving other AMOSR models.

The OSAttack algorithm for CGDL generates adversarial examples by integrating $\ell_{r}$ with $\ell_{s}$ constraints. 
The $\ell_{s}$ modifies the probability distribution of the output layer to make the probabilistic responses of the adversarial examples more similar to those of closed set, 
effectively disguising them as closed set samples and blurring the distinction between open set and closed set. 
Meanwhile, $\ell_{r}$ optimizes the performance of adversarial examples in the encoder and decoder, 
making their intermediate feature representations closer to known samples, 
enhancing their deceptiveness, and improving the transferability and stealth of the attacks. 
This demonstrates that effective attacks can be achieved even without specific knowledge of the AMOSR models, 
and it suggests that future efforts could focus on enhancing transferability through increased constraints.
\section{Summary}\label{sFive}
%
In this paper, we propose OSAttack attack to evaluate the adversarial vulnerability of AMOSR models.
By analyzing the characteristics of AMOSR and constructing an adversarial threat model for AMOSR. 
We delve into the discriminative mechanisms of four typical OSR models and apply the OSFGSM and OSPGD algorithms to generate adversarial examples at the decision level of these models. 
Qualitative and quantitative experimental results show that OSAttack is sufficient to significantly weaken the AMOSR performance even with small perturbations. 
Thus, although AMOSR show certain advantages in dealing with unknown signals, they are still vulnerable to well-designed adversarial signals, a finding that provides an important perspective for the research of future defense strategies.
Future work will focus on developing adversarial attacks with feature layer constraints and exploring new attack strategies against specific targets in open sets.
\section*{Acknowledgment}
This work is supported by the National Natural Science Foundation of China under Grant U23A20271 and 62201172. 

\bibliographystyle{IEEEtran}
\bibliography{cite_SC}

\vspace{12pt}

\end{document}